\documentclass[aps,prd,superscriptaddress,twocolumn]{revtex4-2}
\usepackage[utf8]{inputenc}
\usepackage{dsfont}
\usepackage{amsfonts}
\usepackage{amsmath}
\allowdisplaybreaks[4]        
\usepackage{amssymb}
\usepackage{euscript}     
\usepackage{braket}
\usepackage{starfont}
\usepackage{color,soul}         
\usepackage{tensor}        
\usepackage{amsthm}
\usepackage{graphicx}
\usepackage{slashed}
\usepackage{leftidx}
\usepackage{subfigure}
\usepackage{bbm}
\definecolor{outerspace}{rgb}{0.25, 0.29, 0.3}
\definecolor{scarlet}{rgb}{1.0, 0.13, 0.0}
\usepackage[header,title,page,titletoc]{appendix}  
\definecolor{princetonorange}{rgb}{1.0, 0.56, 0.0}
\definecolor{WildStrawberry}{rgb}{1.0, 0.26, 0.64}
\definecolor{rossocorsa}{rgb}{0.83, 0.0, 0.0}
\definecolor{navyblue}{rgb}{0.0, 0.0, 0.5}
\usepackage{float}
\usepackage[paper=letterpaper,margin=1in]{geometry}

\usepackage{mathtools}

\DeclareMathAlphabet{\pazocal}{OMS}{zplm}{m}{n}
\newcommand{\req}[1]{(\ref{#1})} 

\newcommand{\bea}{\begin{eqnarray}}

\newcommand{\eea}{\end{eqnarray}}
\newcommand{\ba}{\begin{eqnarray}}
\newcommand{\ea}{\end{eqnarray}}

\newcommand{\be}{\begin{equation}}
\newcommand{\ee}{\end{equation} }
\newcommand{\beqa}{\begin{eqnarray}}
\newcommand{\eeqa}{\end{eqnarray}}
\newcommand{\beqar}{\begin{eqnarray*}}
\newcommand{\eeqar}{\end{eqnarray*}}

\renewcommand{\req}[1]{(\ref{#1})}

\newcommand{\eg}{{\it e.g.,}\ }

\makeatletter
\newcommand{\dal}{\mathop{\mathpalette\dal@\relax}}
\newcommand{\dal@}[2]{%
  \begingroup
  \sbox\z@{$\m@th#1\square$}%
  \dimen0=\fontdimen8
    \ifx#1\displaystyle\textfont\else
    \ifx#1\textstyle\textfont\else
    \ifx#1\scriptstyle\scriptfont\else
    \scriptscriptfont\fi\fi\fi3
  \makebox[\wd\z@]{%
    \hbox to \ht\z@{%
      \vrule width \dimen0
      \kern-\dimen0
      \vbox to \ht\z@{
        \hrule height \dimen0 width \ht\z@
        \vss
        \hrule height 2\dimen0
      }%
      \kern-2.5\dimen0
      \vrule width 2.5\dimen0
    }%
  }%
  \endgroup
}
\makeatother

\usepackage{hyperref}
\hypersetup{
    colorlinks,
    citecolor=navyblue,
    filecolor=navyblue,
    linkcolor=navyblue,
    urlcolor=navyblue
}

\begin{document}

\title{Dynamical Formation of Regular Black Holes}
\author{Pablo Bueno}
\email{pablobueno@ub.edu}
\affiliation{Departament de F\'isica Qu\`antica i Astrof\'isica, Institut de Ci\`encies del Cosmos\\
 Universitat de Barcelona, Mart\'i i Franqu\`es 1, E-08028 Barcelona, Spain }

\author{Pablo A. Cano}
\email{pablo.cano@icc.ub.edu}
\affiliation{Departament de F\'isica Qu\`antica i Astrof\'isica, Institut de Ci\`encies del Cosmos\\
 Universitat de Barcelona, Mart\'i i Franqu\`es 1, E-08028 Barcelona, Spain }

\author{Robie A. Hennigar}
\email{robie.a.hennigar@durham.ac.uk}
\affiliation{Departament de F\'isica Qu\`antica i Astrof\'isica, Institut de Ci\`encies del Cosmos\\
 Universitat de Barcelona, Mart\'i i Franqu\`es 1, E-08028 Barcelona, Spain }
\affiliation{Centre for Particle Theory, Department of Mathematical Sciences, Durham University, Durham DH1 3LE, U.K.}

\author{\'Angel J. Murcia}
\email{angelmurcia@icc.ub.edu}
\affiliation{Departament de F\'isica Qu\`antica i Astrof\'isica, Institut de Ci\`encies del Cosmos\\
 Universitat de Barcelona, Mart\'i i Franqu\`es 1, E-08028 Barcelona, Spain }
 \affiliation{INFN, Sezione di Padova, Via Francesco Marzolo 8, I-35131 Padova, Italy}


\begin{abstract}

We study dynamical gravitational collapse in a theory with an infinite tower of higher-derivative corrections to the Einstein-Hilbert action and we show that, under very general conditions, it leads to the formation of regular black holes. Our results are facilitated by the use of a class of  theories that possess second-order equations on spherically symmetric metrics, but which are general enough to provide a basis for the gravitational effective action. 
We analytically solve the collapse of a thin shell of dust and show that it inevitably experiences a bounce at small radius and that its motion can be extended to arbitrary proper time. The collapse of the shell always gives rise to a singularity-free, geodesically complete spacetime that contains horizons if the total mass is above a critical value. In that case, the shell bounces into a new universe through a white hole explosion.  Our construction provides, to the best of our knowlege, the first fully dynamical description of formation of regular black holes, and it suggests that higher-derivative corrections may be the most natural way to resolve the singularities of Einstein's theory.    

\end{abstract}
\maketitle

{\bf Introduction}. According to General Relativity (GR),  the gravitational collapse of ordinary matter leads to the formation of black holes which hide spacetime singularities in their interiors  \cite{Hawking:1973uf, Senovilla:1998oua}. Finding a mechanism for the resolution of such singularities is one of the most prominent open problems in fundamental physics.


One approach entails considering \emph{ad hoc} modifications of known black hole solutions whose singular interiors are thereby replaced by regular cores \cite{Sakharov:1966aja,1968qtr..conf...87B,1981NCimL......161G,Dymnikova:1992ux,Borde:1994ai, Mars:1996khm,Borde:1996df,Ansoldi:2008jw,Hayward:2005gi,Lemos:2011dq,Bambi:2013ufa,Simpson:2018tsi,Rodrigues:2018bdc,Vagnozzi:2022moj,Ovalle:2023ref,Pedrotti:2024znu}. While modifying by hand a singular metric in order to make it regular is  usually a straightforward exercise, finding  regular black holes as \emph{solutions} to actual gravitational theories  is a significantly greater challenge.
For instance, there has been progress in embedding regular black holes as solutions of GR minimally coupled to exotic matter \cite{Ayon-Beato:1998hmi,Bronnikov:2000vy,Ayon-Beato:2000mjt,Bronnikov:2000yz,Ayon-Beato:2004ywd,Dymnikova:2004zc,Berej:2006cc,Balart:2014jia,Fan:2016rih,Bronnikov:2017sgg,Junior:2023ixh,Alencar:2024yvh,Bronnikov:2024izh,Bolokhov:2024sdy,Skvortsova:2024wly,Murk:2024nod,Li:2024rbw,Zhang:2024ljd}, but this approach is not satisfying as these theories also contain singular solutions --- in fact, these theories contain all the vacuum solutions of GR.\footnote{See 
\cite{Frolov:1989pf,Barrabes:1995nk,Nicolini:2005vd,Olmo:2012nx,Balakin:2015gpq,Bazeia:2015uia,Chamseddine:2016ktu,Bambi:2016xme,Bejarano:2017fgz,Colleaux:2017ibe,Cano:2018aod,Colleaux:2019ckh,Guerrero:2020uhn,Brandenberger:2021jqs,Olmo:2022cui,Junior:2024xmm,Frolov:2014jva} for alternative scenarios. }

A proper resolution of singularities  must therefore involve a modification of gravitational dynamics. This aligns with the idea that GR is not a complete theory and that it must be modified in large curvature regimes. In particular, it is expected that these modifications take the form of higher-curvature corrections to the Einstein-Hilbert action --- see \eg~\cite{Gross:1986iv,Gross:1986mw,Metsaev:1987zx,Bergshoeff:1989de}. In this context, it was first shown that the singularities of charged black holes can be resolved by higher-derivative terms with nonminimal couplings \cite{Cano:2020qhy,Cano:2020ezi,Bueno:2021krl}, but this requires having a nonzero charge. The existence of regular black holes from pure gravity remained elusive until recently,  when some of us \cite{Bueno:2024dgm} showed that the Schwarzschild singularity in $D\geq 5$ spacetime dimensions gets fully resolved by supplementing the Einstein-Hilbert action by infinite towers of higher-curvature corrections.\footnote{Various direct follow-ups to this paper have appeared since then, including \cite{Konoplya:2024hfg, DiFilippo:2024mwm, Konoplya:2024kih, Ma:2024olw, Ditta:2024iky, Frolov:2024hhe}.} This is achieved generically, without any fine-tuning amongst the gravitational couplings, provided they satisfy certain mild constraints. 
The models involve densities of arbitrarily high curvature order that belong to the class of ``Quasi-topological gravities'' \cite{Oliva:2010eb,Quasi,Dehghani:2011vu,Ahmed:2017jod,Cisterna:2017umf,Bueno:2019ycr, Bueno:2022res, Moreno:2023rfl,Moreno:2023arp}, and are broad enough to provide a basis for the gravitational effective action \cite{Bueno:2019ltp,Bueno:2024dgm}. Hence, even though the singularity resolution requires going beyond the perturbative regime of the gravitational couplings, the hope is that the result captures some features of a full quantum theory of gravity.

Some of the most important open questions of regular black holes concern their dynamical aspects, such as their formation and stability \cite{Hayward:2005gi,Carballo-Rubio:2018pmi,Carballo-Rubio:2021bpr,DiFilippo:2022qkl, Carballo-Rubio:2022kad}. So far, these questions have not been studied with enough rigor due to the lack of a dynamical theory that predicts regular black holes. 
The goal of this letter is to show that the theories of \cite{Bueno:2024dgm} not only predict regular black holes, but that they provide a full dynamical description of gravitational collapse leading to the formation of such black holes. 
To this end, we show that these theories give rise to stable time evolution within spherical symmetry, and we solve explicitly the problem of thin-shell collapse.  In a companion paper~\cite{collapseLong} we provide additional details and further extend the results reported here.

{\bf Quasi-topological gravities}. 
From a bottom-up perspective, a gravitational effective action can be built by including all possible diffeomorphism-invariant terms in a perturbative expansion controlled by (a priori) unconstrained couplings. Such terms can be modified by perturbative field redefinitions of the metric and hence different bases of invariants may be chosen. 
In this letter we consider a particular basis of densities\footnote{The fact that those densities provide a basis for the gravitational effective action follows from an argument which is essentially identical to the one presented in the appendix of \cite{Bueno:2024dgm}.} which exists in $D\geq 5$ and whose action can be written as
\begin{equation}\label{QTaction}
S_{\rm QT}= \int \frac{\mathrm{d}^Dx \sqrt{|g|}}{16\pi G_{\rm N}}  \left[R+\sum_{n=2}^{n_{\rm max}} \alpha_n \mathcal{Z}_n \right]\, ,
\end{equation}
where  $G_{\rm N}$ is the Newton constant and 
$\alpha_n$ are arbitrary coupling  constants with dimensions of  length$^{2(n-1)}$. The densities $\mathcal{Z}_n$ are selected by the condition that they possess second-order equations on general spherically symmetric (SS) ans\"atze. In particular, they belong to a broader family of theories known as Quasi-topological (QT) gravities \cite{Oliva:2010eb,Quasi,Dehghani:2011vu,Ahmed:2017jod,Cisterna:2017umf,Bueno:2019ltp,Bueno:2019ycr, Bueno:2022res, Moreno:2023rfl,Moreno:2023arp}. The densities $\mathcal{Z}_n$ for $n=2,3,4,5$ can be found in appendix \ref{QTGs} and it is convenient to define $\mathcal{Z}_{1}\equiv R$. Arbitrarily higher-order densities can be obtained from the following recursive formula  \cite{Bueno:2019ycr} 
 \begin{align}\notag
\mathcal{Z}_{n+5}=
&+\frac{3(n+3)\mathcal{Z}_{1}\mathcal{Z}_{n+4}}{D(D-1)(n+1)}-\frac{3(n+4)\mathcal{Z}_{2}\mathcal{Z}_{n+3}}{D(D-1)n}\\ \label{zrec}&+\frac{(n+3)(n+4)\mathcal{Z}_{3}\mathcal{Z}_{n+2}}{D(D-1)n(n+1)}\, .
\end{align}
When the seed densities possess second-order equations on general SS metrics, the recursive formula preserves this property --- see the appendix and \cite{collapseLong}.


 {\bf Effective Two-dimensional Action}. 
To study the spherically symmetric equations of motion of  \req{QTaction}, it is useful to dimensionally reduce it on a $(D-2)$-sphere. 
Thus, we evaluate the  QT action on a metric of the form
\begin{equation}
\mathrm{d}s^2=\gamma_{\mu\nu}\mathrm{d} x^{\mu}\mathrm{d}x^{\nu}+\varphi(x)^2 \mathrm{d}\Omega^2_{D-2}\, ,
\end{equation}
where $\mathrm{d}\Omega^2_{D-2}$ is the sphere metric.
After a long calculation, we obtain a reduced action for the two-dimensional metric $\gamma_{\mu\nu}$ and for the radial scalar $\varphi$, 
\begin{equation}\label{2dtheory}
S_{\rm 2d}= \frac{(D-2) \Omega_{(D-2)}}{16 \pi G_{\rm N}} \int \mathrm{d}^{2}x\sqrt{|\gamma|} \mathcal{L}_{\rm 2d}(\gamma_{\mu\nu},\varphi) \, ,
\end{equation}
where 
$\Omega_{(D-2)}\equiv 2\pi^{(D-1)/2}/\Gamma[\tfrac{D-1}{2}]$
is the sphere volume. The key observation is that, since by construction \req{QTaction} yields second-order SS equations, then this action must be a Horndeski theory \cite{Horndeski:1974wa}. This expectation is borne out --- we find that the Lagrangian takes the Horndeski form,
\begin{align}\notag 
&\mathcal{L}_{\rm 2d}=G_{2}(\varphi, X)-\Box\varphi G_{3}(\varphi, X)+G_{4}(\varphi, X)R\\  \label{horn}&-2G_{4,X}(\varphi, X)\left[(\Box\varphi)^2-\nabla_{\mu}\nabla_{\nu}\varphi\nabla^{\mu}\nabla^{\nu}\varphi\right]\, ,
\end{align}
where $X\equiv \nabla_{\mu}\varphi\nabla^{\mu}\varphi$, $G_{4,X}\equiv \partial_{X}G_{4}$, and 
\begin{align}
G_{2}(\varphi, X)&=\varphi^{D-2}\left[(D-1)h(\psi)-2\psi  h'(\psi)\right]\, ,\\\label{G3form}
G_{3}(\varphi, X)&=2\varphi^{D-3}h'(\psi)\, ,\\
\label{eq:G4}
G_{4}(\varphi, X)&=-\frac{\varphi^{D-2}}{2}\psi^{(D-2)/2}\int \mathrm{d}\psi\frac{h'(\psi)}{ \psi^{D/2}}\, ,
\end{align}
and where we defined 
 \be \label{eom_psi}
h(\psi)\equiv \psi + \sum_{n=2}^{n_{\rm max}}\alpha_n \frac{(D-2n)}{(D-2)}  \psi^n\, ,\quad  \psi \equiv  \, \frac{1-X}{\varphi^2}  \, .
\ee
The \textit{characteristic polynomial} $h(\psi)$ is an useful object which encapsulates many features of QT gravity solutions --- see \eg \cite{Bueno:2022res,Bueno:2020odt,Camanho:2011rj}. 


 
 {\bf Birkhoff Theorem.}
 The variation of \req{2dtheory} with respect to $\gamma_{\mu\nu}$ and $\varphi$ yields the spherically symmetric equations of motion of the higher-dimensional theory  \req{QTaction}, $\mathcal{E}_{ab}=0$. 
If we consider the ansatz
\begin{equation}
\mathrm{d}s_{\gamma}^2=-N(t,r)^2 f(t,r) \mathrm{d}t^2+\frac{\mathrm{d}r^2}{f(t,r)} 
\label{eq:ssans}
\end{equation}
for $\gamma_{\mu\nu}$, 
and set $\varphi=r$ (which implies $X=f(t,r)$), the $(t,r)$ components of the equations read 
\begin{align}\label{Ett}
\mathcal{E}_{tt}&=\frac{(D-2) N^2 f}{2r^{D-2}} \frac{\partial }{\partial r} \left[r^{D-1}h(\psi) \right] \,,\\\label{Etr}
\mathcal{E}_{tr}&=-\frac{(D-2) \partial_t f}{2 r f}  h'(\psi) \,  \,,\\\label{Err}
\mathcal{E}_{rr}&=\frac{(D-2) \partial_r N }{r N} h'(\psi)-\frac{1}{N^2 f^2}\mathcal{E}_{tt} \,.
\end{align}
These come from the variation of \req{2dtheory} with respect to $\gamma_{\mu\nu}$. The variation with respect to $\varphi$ yields the angular components of the higher-dimensional equations of motion, which are related to the $(t,r)$ ones via Bianchi identities. 

Observe that the equations \req{Ett}-\req{Err} are of first order and only differ from those of GR via the function $h(\psi)$. 
 Now, it is straightforward to verify that imposing $\mathcal{E}_{ab}=0$ leads to the conditions
 \begin{equation}
 \partial_t f=0\, , \quad \partial_r N=0\, , \quad  \frac{ \partial}{{\partial} r} \left[r^{D-1}h(\psi) \right]=0\, .
 \label{eq:eomhorn}
 \end{equation}
 Hence, $f=f(r)$ and $N=N(t)$, which can be reabsorbed in a redefinition of the time coordinate $N(t)^2 {\mathrm d}t^2 \rightarrow  {\mathrm d}t^2$. We thus conclude that the most general spherically symmetric solution of \req{QTaction} is in fact static and fully determined by a single function
 \begin{equation}
\mathrm{d}s^2=- f(r) \mathrm{d}t^2+\frac{\mathrm{d}r^2}{f(r)}+r^2 \mathrm{d}\Omega^2_{D-2}\,.
\label{eq:ssans}
\end{equation}
The metric function $f(r)$ is uniquely determined by the algebraic equation
 \be \label{eom}
h(\psi)  = \frac{2\mathsf{M}}{r^{D-1}}  \, ,\quad \psi=\frac{1-f(r)}{r^2}\, ,
\ee
where  $\mathsf{M}$ is an integration constant related to the ADM mass \cite{Arnowitt:1960es,Arnowitt:1960zzc,Arnowitt:1961zz,Deser:2002jk} of the solution, $M$, through 
\begin{equation} \label{simpleM}
\mathsf{M} \equiv \frac{8\pi G M}{(D-2)\Omega_{(D-2)}}\, .
\end{equation}
This proves that a Birkhoff theorem is satisfied for QT theories of arbitrarily high curvature orders and in general dimensions $D\geq 5$, extending previous partial results presented in \cite{Oliva:2010eb, Oliva:2011xu, Cisterna:2017umf}. 

{\bf Regular Black Holes}. 
As \cite{Bueno:2024dgm} realized, when we consider an infinite tower of corrections, $n_{\rm max}\to\infty$, the solutions of \req{eom} are singularity-free under very general conditions. For instance, the conditions $
\alpha_{n}(D-2n)\ge 0\,\, \forall\, n$,  $\lim_{n\rightarrow\infty} |\alpha_{n}|^{\frac{1}{n}}=C>0$ are sufficient to ensure regularity. 

The explicit form of these regular black holes can be obtained for specific choices of $\alpha_{n}$ \cite{Bueno:2024dgm,DiFilippo:2024mwm,Konoplya:2024kih}. To illustrate our results, we will consider the example $\alpha_n (D-2n)=(D-2)\alpha^{n-1}$.\footnote{We are implicitly assuming that $D$ is odd. When $D$ is even we have $\alpha_{n}(D-2n)=0$ for $n=D/2$, which makes it more difficult to find simple summable examples of $h(\psi)$ \req{eom_psi}. However, this is merely a practical problem --- see \cite{collapseLong} for more details.} In this case the series \req{eom_psi} yields $h(\psi)=\psi/(1-\alpha\psi)$, and the metric function $f(r)$ takes the form
\begin{equation}\label{hayf}
f(r)=1-\frac{2\mathsf{M} r^2}{r^{D-1}+2\mathsf{M} \alpha }\, ,
\end{equation}
which is the $D$-dimensional Hayward black hole. Define the critical mass $\mathsf{M}_{\rm cr}=2\alpha$. When $\mathsf{M}>\mathsf{M}_{\rm cr}$, this spacetime has an outer and an inner horizon. In $D=5$ these are located at
\begin{equation}
r_{\pm}=\sqrt{\mathsf{M} \pm \sqrt{\mathsf{M} (\mathsf{M}-2\alpha)}}\, .
\end{equation}
For $\mathsf{M}=\mathsf{M}_{\rm cr}$, the regular black hole is extremal and below the critical mass, the solution is a gravitating soliton.
Naturally, this solution reduces to the usual Schwarzschild-Tangherlini solution, 
\begin{equation}\label{Sch}
f(r)=1-\frac{2\mathsf{M} }{r^{D-3}}\, ,
\end{equation}
for $\alpha=0$ as well as for large radius. For small $r$ we have $f(r)\approx 1-r^2/\alpha$ and the singularity is replaced by a regular core as long as $\alpha>0$.

\begin{figure*}
\centering \hspace{0cm}
\includegraphics[width=0.487\textwidth]{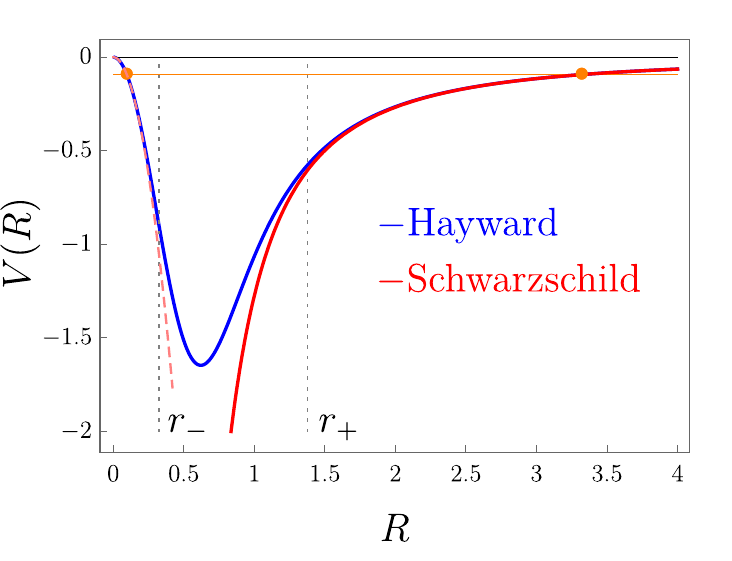}
\includegraphics[width=0.501\textwidth]{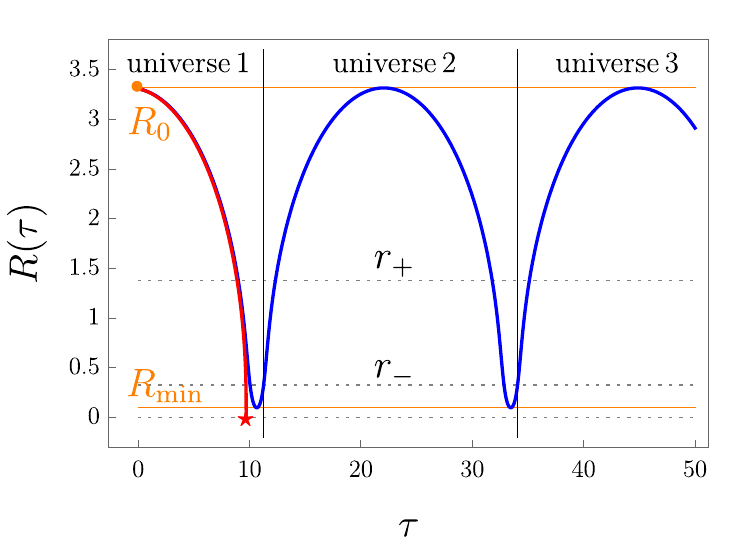}
\caption{(Left) Effective potential $V(R)$ of the thin-shell equation \req{tse} in the case of $D=5$ Einstein gravity (red) and a QT gravity theory of the form \req{QTaction} with $\alpha_n=\tfrac{3}{(5-2n)}\alpha^{n-1}$ (blue). We have set $\mathsf{M}=1$, $\mathsf{m}=1.05$ and $\alpha=1/10$. 
The dashed pink line shows the analytical approximation \req{approx} for the effective potential near the origin and the orange points are the turning points of the second potential. 
(Right) Shell radius as a function of the proper time $\tau$.
 For Einstein gravity, the shell collapses reaching $R=0$ after a finite proper time (red star), forming a Schwarzschild black hole. For the QT theory, the shell collapses forming a Hayward black hole, it reaches some finite minimum radius $R_{\rm min}$ and bounces back, emerging in a new universe, where the process is repeated once it reaches $R_0$ again. }
\label{fig:hay}
\end{figure*}

{\bf Thin-shell collapse}. Let us now consider the collapse of a thin spherical shell of pressureless matter (``dust''). The surface stress-energy tensor takes the form $S_{AB} = \sigma u_A u_B$, where $\sigma$ is the surface energy density of the matter and $u_A$ is its $D$-velocity. In a proper time (denoted by $\tau$) parametrization of the shell, the components of the surface stress-energy tensor are simply
\be 
S_{\tau\tau} = \sigma \, , \quad S_{ij} = 0 \, ,
\ee
where $i,j$ are the angular components. 
At a given moment of proper time, we set the radius of the shell to $r = R(\tau)$. Inside the shell, $r < R(\tau)$,  we take the metric to be Minkowski space. By Birkhoff's theorem, the exterior of the shell, $r > R(\tau)$, is necessarily the unique solution of~\eqref{eom}. Therefore, the spacetime metric consists of two charts that are joined at the location of the shell,
\be 
{\rm d} s_\pm^2 = - f_\pm (r)  {\rm d}t_\pm^2 + \frac{ {\rm d}r ^2}{f_\pm(r)} + r^2 {\rm d} \Omega^2_{D-2} \, ,
\ee
where $f_-(r) = 1$ corresponds to the inner Minkowski region and $f_+(r)$ is the solution of~\eqref{eom}. 

The spacetime trajectory of the shell, $(t_\pm, r) = \left(T_\pm(\tau), R(\tau) \right)$, is determined by the junction conditions appropriate to the theory. The first junction condition, as in GR, requires that the metric be continuous across the shell. The induced metric on the shell is 
\be 
{\rm d}s_\Sigma^2 = - \left(f_\pm(R) \dot{T}_\pm^2 - \frac{\dot{R}^2}{f_\pm(R)} \right) d\tau^2 + R(\tau)^2 {\rm d} \Omega_{D-2}^2 \, ,
\ee
and demanding continuity (taking into account that $\tau$ is the proper time), we find that
\be 
f_{\pm}(R)\dot{T}_\pm = \sqrt{f_\pm(R) + \dot{R}^2} \equiv \beta_\pm \,  .
\ee

The second junction condition is more subtle. It is most simply obtained from the action principle as the boundary equations of motion~\cite{Brown:1992br, Davis:2002gn},
\be 
\Pi_{AB}^- - \Pi_{AB}^+ = 8 \pi G_{\rm N} S_{AB} \, , \quad \Pi_{AB} \equiv \frac{16 \pi G_{\rm N}}{\sqrt{|h|}} \frac{\delta S^{\rm total}}{\delta h^{AB}} \, ,
\ee
where $h_{AB}$ is the boundary metric and  $S^{\rm total}=S+S^{\rm bdry}$ is the total gravitational action including boundary terms that make the variational principle well-posed. As we explain in \cite{collapseLong}, in spherical symmetry the computation of $\Pi_{AB}$ can be rigorously performed with the aid of the two-dimensional action \req{2dtheory}, whose boundary terms are known \cite{Padilla:2012ze}. The final result reads
\begin{align}
\Pi_{\tau\tau}^- - \Pi_{\tau\tau}^+ &= 8 \pi G_{\rm N} \sigma \, ,
\\
\frac{{\rm d}}{{\rm d} \tau} \left[R^{D-2}\left(\Pi_{\tau\tau}^- - \Pi_{\tau\tau}^+ \right) \right] &= 0 \, ,
\end{align}
where 
\be 
\Pi_{\tau\tau}^\pm = \frac{(D-2)}{R} \int_0^{\beta_\pm} {\rm d}z  h' \left(\frac{1 + \dot{R}^2 - z^2}{R^2} \right) \, .
\ee
The second of the two equations above implies that the shell's proper mass is conserved, 
\be 
m \equiv \sigma \Omega_{D-2} R^{D-2}  = {\rm constant} \, .
\ee
On the other hand, the first equation can be reduced to a master equation that determines the motion of the shell, 
\begin{equation}\label{sdd}
\frac{\mathsf{m}}{R^{D-1}}=\int_{R}^{\infty} \frac{\mathrm{d}r \, \mathsf{M} (D-1)}{r^D \sqrt{1+\dot R^2-\frac{R^2}{r^2}\left[1-f(r) \right]}}\, ,
\end{equation}
where we defined $\mathsf{m} \equiv 8 \pi G_{\rm N} m/[(D-2) \Omega_{D-2}]$ in analogy with~\eqref{simpleM} to ease notation. This is one of the main results of our study. It is valid for any QT theory of the class considered in \req{QTaction}. Plugging the black hole metric function $f(r)$ for the chosen model yields an integro-differential equation for $R(\tau)$, the solution of which determines the fate of the collapsing shells.

It is helpful to recast~\eqref{sdd} in the form 
\begin{equation}\label{tse}
\dot R^2+V(R)= \frac{\mathsf{M}^2}{\mathsf{m}^2}-1 \, , 
\end{equation}
where $V(R)$ is an effective potential. In the case of Einstein Gravity, for which $f(r)$ is given by \req{Sch}, the potential can be easily found to be
\begin{equation}\label{VGR}
V(R)=-\frac{\mathsf{M} }{R^{(D-3)}}-\frac{\mathsf{m}^2}{4R^{2(D-3)}}\, .
\end{equation}
This is a monotonously decreasing function of $R$, and starting at any finite radius $R(0)=R_0$, the shell collapses leaving behind a Schwarzschild black hole and reaching $R=0$ after a finite proper time --- see Fig.\,\ref{fig:hay}.

For any QT theory admitting regular black hole solutions and a Birkhoff theorem, since the metric function is asymptotically given by the Schwarzschild metric to leading order, the large $R$ behaviour of the potential is the same as in Einstein gravity~\eqref{VGR}. On the other hand, the small $R$ behaviour of the metric is completely altered. Using the fact that the metric function of a regular black hole behaves near $r = 0$ as
\be 
f(r) = 1 - \frac{r^2}{C} + \cdots
\ee
for some constant $C$, we obtain for the effective potential
\be \label{approx}
V(R) = - \frac{R^2}{C} \quad {\rm for} \quad R \to 0 \, .
\ee
Rather than diverge, the potential limits to zero as $R \to 0$ and is smooth. 
Constructing the potential at intermediate values of $R$ generally requires numerical methods.  We show this result for the $D = 5$ Hayward black hole --- see \req{hayf} --- in Fig.\,\ref{fig:hay}. While this plot is for a \textit{particular} solution, we observe that the qualitative features are \textit{generic}, \textit{i.e.}, it vanishes at the origin and at infinity and it contains a minimum at some intermediate $R$. 

\begin{figure}[t!]
 \centering
              \includegraphics[scale=0.47,trim={0 0.18cm 0 0},clip]{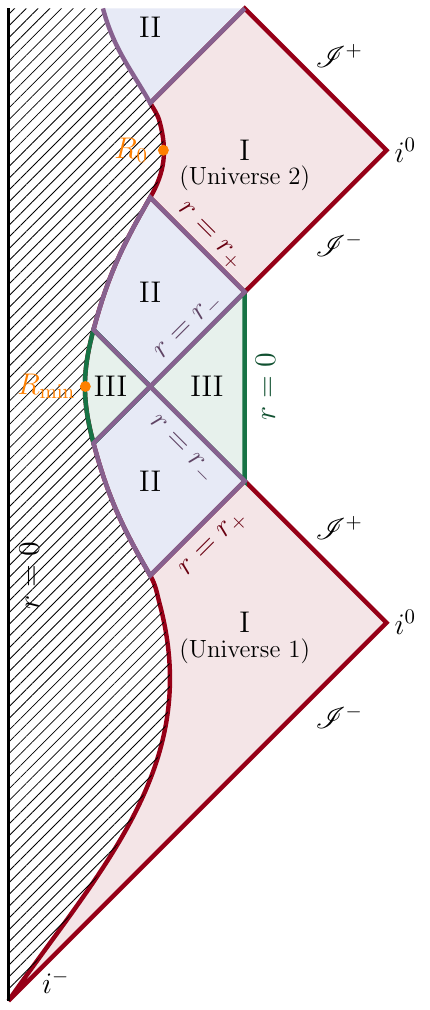}           
              \caption{Penrose diagram associated with the dynamical collapse of a spherical thin shell in $D=5$ in the theory \eqref{QTaction} with $(5-2n)\alpha_n=3 \alpha^{n-1}$. It is assumed that matter existed from ancient times and was assembled into a spherical thin shell at some point by a future civilization.}
              \label{fig:pd}
          \end{figure}

If the total mass is above the critical threshold $\mathsf{M}>\mathsf{M}_{\rm cr}$, a shell that starts collapsing at some finite radius  $R_0>r_+$ keeps on decreasing its size, eventually giving rise to a regular black hole. As the shell continues to collapse, it will cross $r=r_-$ and the inner horizon will form. By the use of Kruskal-Szekeres-like coordinates, the original coordinate patch may be extended into two causally disconnected regions $r_- \geq r \geq 0$ (regions III in Fig. \ref{fig:pd}). One of these contains the shell, inside of which the spacetime is flat, while the other region corresponds to a  ``de Sitter core'' in which the line $r=0$ is fully regular.
\footnote{The diagram in Fig.~\ref{fig:pd} resembles that of the collapse of a charged spherical shell in Einstein-Maxwell theory, but with the singularity removed, see \eg \cite{IsraelBounce}.} Ultimately, the shell starts climbing the potential and reaches a turning point at which $\dot{R} = 0$ and $R = R_{\rm min}$ --- this always happens in region III. At that point, a bounce occurs. The shell begins increasing its size, crossing the inner and outer horizons and emerging in a new universe from a white hole. The shell will grow up to $r=R_0$, at which point the process of collapse restarts. If the total mass is below the critical threshold, the shell experiences a bounce as well, but horizons never form.

{\bf Discussion}. We have shown that regular black holes are the endpoint of gravitational collapse in a purely gravitational theory. This is a \textit{generic consequence} of a theory containing an infinite tower of higher-curvature corrections with only very mild and qualitative conditions on the couplings. This provides a mechanism for the resolution of singularities and the formation of regular black holes in any dimension $D \ge 5$. We believe this is the first time results of such generality have been achieved.

Our model affords considerable opportunity to address important problems in the theory of regular black holes and singularity resolution. Among these, for example, is the possibility to consider more complicated shell configurations~\cite{Cardoso:2016wcr}, other forms of matter collapse, or to consider the problem of critical scaling~\cite{Choptuik:1992jv}. The stability problem of the inner horizon can also be studied in the spherically symmetric sector. Moreover, the two-dimensional Horndeski theory we have identified can be utilized to understand the effects of strong quantum gravitational fluctuations in the vicinity of near extremal regular black holes~\cite{Iliesiu:2020qvm}. These effects will likely play an important role in the final stages of regular black hole evaporation. 

Whether or not the mechanism we have identified is the one responsible for singularity resolution in Nature remains to be seen. What is clear is that it provides a robust mechanism where many long-thought impossible questions can be finally addressed.

\vspace{0.1cm}
\begin{acknowledgments} 
We would like to thank Javier Moreno, Simon Ross, and Guido van der Velde for useful conversations. PB was supported by a Ramón y Cajal fellowship (RYC2020-028756-I), by a Proyecto de Consolidación Investigadora (CNS 2023-143822) from Spain’s Ministry of Science, Innovation and Universities, and by the grant PID2022-136224NB-C22, funded by MCIN/AEI/ 10.13039/501100011033/FEDER, UE.
The work of PAC received the support of a fellowship from “la Caixa” Foundation (ID 100010434) with code LCF/BQ/PI23/11970032. 
The work of RAH received the support of a fellowship from ``la Caixa” Foundation (ID 100010434) and from the European Union’s Horizon 2020 research and innovation programme under the Marie Skłodowska-Curie grant agreement No 847648 under fellowship code LCF/BQ/PI21/11830027. {\'A}JM was supported by a Juan de la Cierva contract (JDC2023-050770-I) from Spain’s Ministry of Science, Innovation and Universities. {\'A}JM would like to thank the University of Barcelona for its warm hospitality before the start of the contract.
\end{acknowledgments}

\onecolumngrid  
\begin{center}  
{\Large\bf Appendices} 
\end{center} 
\appendix 

\section{Quasi-topological gravities satisfying a Birkhoff theorem}\label{QTGs}

Let $\mathcal{L}(g^{ab},R_{cdef})$ be any $D$-dimensional (with $D \geq 5$) higher-curvature theory of gravity constructed from arbitrary contractions of the Riemann curvature tensor with the metric. In particular, no covariant derivatives of the curvature are assumed to appear. The gravitational equations of motion of $\mathcal{L}(g^{ab},R_{cdef})$ take the following form \cite{Padmanabhan:2011ex}:
\begin{equation}
P_{acde} R_{b}{}^{cde}-\frac{1}{2}\mathcal{L}g_{ab}+2\nabla^c \nabla^d P_{acbd}=0\,,
\end{equation}
where we defined $P^{abcd}=\frac{\partial \mathcal{L}}{\partial R_{abcd}}$. Take a general $D$-dimensional spherically symmetric ansatz for the metric (the arguments also work for planar or hyperbolic symmetry):
\begin{equation}
\mathrm{d}s_{N,f}^2=-N(t,r)^2 f(t,r) \mathrm{d}t^2+\frac{1}{f(t,r)} \mathrm{d}r^2+r^2 \mathrm{d} \Omega^2_{D-2}\,,
\label{eq:appss}
\end{equation}
where $\mathrm{d}\Omega^2_{D-2}$ stands for the metric of the round $(D-2)$-dimensional sphere. In this manuscript we have considered those higher-curvature theories of gravity $\mathcal{L}(g^{ab},R_{cdef})$ for which the equations of motion on top of \eqref{eq:appss} are strictly of second order in derivatives and, furthermore, satisfy naturally a Birkhoff theorem. Specifically, we require that
\begin{equation}
\nabla^c \nabla^d P_{acbd} \vert_{N,f}=0\,,
\label{eq:appc1}
\end{equation}
where $\vert_{N,f}$ denotes evaluation on \eqref{eq:appss}, and that spherical symmetry at the level of the equations of motion further implies the staticity of the solution. Concretely, this will happen if the equations of motion demand that 
\begin{equation}
\partial_r N=\partial_t f=0\,.
\label{eq:appc2}
\end{equation}
Therefore, one can always set $N=1$ after a time reparametrization, if needed. As it turns out, higher-curvature theories fulfilling this condition may be found at all curvature orders. They correspond to a special subclass of the set of Quasi-topological gravities, characterized by admitting non-hairy generalizations of the static Schwarzschild-Tangherlini solution with $N=1$ (see \cite{Oliva:2010eb,Quasi} and further bibliography cited in the main text). 

Let $\mathcal{Z}_{(n)}$ denote a $D$-dimensional (with $D \geq 5$) theory of gravity constructed from $n$-th order curvature invariants fulfilling conditions \eqref{eq:appc1} and \eqref{eq:appc2}. If $W_{abcd}$ denotes the Weyl curvature tensor and $Z_{ab}$ the traceless part of the Ricci curvature tensor, instances of such theories  up to $n=5$ read as follows:
\begin{subequations}\label{Znexplicit}
\begin{align}
 \mathcal{Z}_{(1)}&=R\,, \\
 \mathcal{Z}_{(2)}&=\frac{1}{(D-2)} \left [\frac{W_{abcd} W^{abcd}}{D-3} -\frac{4 Z_{ab}Z^{ab}}{D-2}\right]+\frac{\mathcal{Z}_{(1)}^2}{D(D-1)}\,, \\
\nonumber \mathcal{Z}_{(3)}&=\frac{24}{(D-2)(D-3)} \left[\frac{ W\indices{_a_c^b^d} Z^a_b Z^c_d}{(D-2)^2}-
   \frac{   W_{a c d e}W^{bcde}Z^a_b}{(D-2) (D-4)}+\frac{2(D-3)
   Z^a_b Z^b_cZ_a^c}{3(D-2)^3} +\frac{(2 D-3) W\indices{^a^b_c_d}W\indices{^c^d_e_f}W\indices{^e^f_a_b}}{12 (D ((D-9)
   D+26)-22)} \right]\\&+ \frac{3\mathcal{Z}_{(1)}\mathcal{Z}_{(2)}}{D(D-1)}-\frac{2 \mathcal{Z}_{(1)}^3}{D^2(D-1)^2} \,, \\
\nonumber
\mathcal{Z}_{(4)}&=\frac{96}{(D-2)^2(D-3)} \left[\frac{(D-1)\left ( W_{abcd} W^{abcd} \right)^2}{8D(D-2)^2(D-3)}-\frac{(2D-3) Z_e^f Z^e_f W_{abcd} W^{abcd}}{4(D-1)(D-2)^2}-
\frac{2 W_{acbd} W^{c efg} W^d{}_{efg} Z^{ab} }{D(D-3)(D-4)}\right. \\ \nonumber & -\frac{4Z_{a c} Z_{d e} W^{bdce} Z^{a}_b}{(D-2)^2(D-4)} \left. +\frac{(D^2-3D+3) \left (Z_a^b Z_b^a\right )^2}{D(D-1)(D-2)^3}-\frac{Z_a^b Z_b^c Z_c^d Z_d^a}{(D-2)^3}+\frac{(2D-1)W_{abcd} W^{aecf} Z^{bd} Z_{ef}}{D(D-2)(D-3)}\right]\\&+\frac{4\mathcal{Z}_{(1)}\mathcal{Z}_{(3)}-3 \mathcal{Z}_{(2)}^2}{D(D-1)}\,,\\ \nonumber
\mathcal{Z}_{(5)}&=\frac{960 (D-1)}{(D-2)^4(D-3)^2} \left[ \frac{(D-2)W_{ghij} W^{ghij}W\indices{_a_b^c^d}W\indices{_c_d^e^f}W\indices{_e_f^a^b} }{40D(D^3-9 D^2+26D-22)}+\frac{4(D-3) Z_a^b Z_b^c Z_c^d Z_d^e Z_e^a}{5(D-1)(D-2)^2(D-4)}\right. \\ \nonumber & -\frac{(3D
-1)W^{ghij} W_{ghij}  W_{a c d e}W^{bcde} Z^a_b}{10D(D-1)^2(D-4)}-\frac{4(D-3)(D^2-2D+2)Z_a^b Z_b^a Z_c^d Z_d^e Z_e^c}{5D(D-1)^2(D-2)^2(D-4)} \\ \nonumber & -\frac{(D-3)(3D-1)(D^2+2D-4)W^{ghij} W_{ghij} Z_c^d Z_d^e Z_e^c}{10D(D-1)^2(D+1)(D-2)^2(D-4)}+\frac{(5D^2-7D+6)Z_g^h Z_h^g W_{abcd} Z^{ac} Z^{bd}}{10D(D-1)^2(D-2)}\\ \nonumber & +\frac{(D-2)(D-3)(15D^5-148 D^4+527 D^3-800 D^2+472D-88)W\indices{_a_b^c^d}W\indices{_c_d^e^f}W\indices{_e_f^a^b} Z_{g}^h Z_h^g}{40D(D-1)^2(D-4)(D^5-15D^4+91 D^3-277 D^2+418D-242)}\\ \nonumber &- \frac{2(3D-1)Z^{ab} W_{acbd} Z^{ef}  W\indices{_e^c_f^g} Z^d_g}{D(D^2-1)(D-4)}-\frac{Z_{a}^b Z_{b}^{c} Z_{cd} Z_{ef} W^{eafd}}{(D-1)(D-2)} +\frac{(D-3)W_{a c d e}W^{bcde} Z^a_b Z_f^g Z_g^f}{5D(D-1)^2(D-4)}\\ \nonumber &\left. -\frac{(D-2)(D-3)(3D-2) Z^a_b Z^b_c W_{daef} W^{efgh} W_{gh}{}^{dc}}{4(D-1)^2(D-4)(D^2-6D+11)}+\frac{W_{ghij}W^{ghij} Z^{ac}Z^{bd}W_{abcd}}{20D(D-1)^2}\right]\\&+\frac{5\mathcal{Z}_{(1)}\mathcal{Z}_{(4)}-2\mathcal{Z}_{(2)}\mathcal{Z}_{(3)}}{D(D-1)}+\frac{6 \mathcal{Z}_{(1)}\mathcal{Z}_{(2)}^2-8 \mathcal{Z}_{(1)}^2\mathcal{Z}_{(3)}}{D^2(D-1)^2}\,.
\end{align}
\end{subequations}
For higher orders of $n$, one may use the recursive formula presented in the main text:
 \begin{align}\label{eq:appzrec}
\mathcal{Z}_{(n+5)}=\frac{3(n+3)\mathcal{Z}_{(1)}\mathcal{Z}_{(n+4)}}{D(D-1)(n+1)}-\frac{3(n+4)\mathcal{Z}_{(2)}\mathcal{Z}_{(n+3)}}{D(D-1)n}+\frac{(n+3)(n+4)\mathcal{Z}_{(3)}\mathcal{Z}_{(n+2)}}{D(D-1)n(n+1)}\, .
\end{align}
The proof that this formula is guaranteed to produce theories satisfying \eqref{eq:appc1} and \eqref{eq:appc2} for arbitrary $n$ goes as follows. First, we note that the first five densities \eqref{Znexplicit} on top of \eqref{eq:appss} may be equivalently written as a Horndeski theory for 2-dimensional gravity with a scalar, as explained in the main text (cf. \eqref{horn}) and proved in the companion paper \cite{collapseLong}. Then, by direct substitution in \eqref{eq:appzrec}, one checks that the equivalent Horndeski theories \eqref{horn} fulfill the recursive relation \eqref{eq:appzrec} for any $n$. Since such Horndeski theories have second order equations and satisfy a Birkhoff theorem (see \eqref{eq:eomhorn}), the $D$-dimensional theories of gravity $\mathcal{Z}_{(n+5)}$ will also comply to conditions \eqref{eq:appc1} and \eqref{eq:appc2} for arbitrary $n$ and we conclude. 

\twocolumngrid 

\bibliographystyle{JHEP-2}
\bibliography{Gravities}
\noindent 


\end{document}
